\documentstyle[epsf]{l-aa} 

\newlength{\fsize}
\begin{document}

\thesaurus{11.03.1 --- 12.04.1 --- 12.07.1 --- 12.12.1} 

\title{Arc statistics with realistic cluster potentials} 
\subtitle{IV. Clusters in different cosmologies} 


\author{Matthias Bartelmann\inst{1} \and Andreas Huss\inst{1} 
\and J\"org M.\ Colberg\inst{1} \and Adrian Jenkins\inst{2} 
\and Frazer R.\ Pearce\inst{2}} 

\offprints{Matthias Bartelmann} 

\institute{Max-Planck-Institut f\"ur Astrophysik, P.O.\ Box 1523, 
D-85740 Garching, Germany 
\and Physics Dept., University of Durham, Durham DH1 3LE, UK} 


\date{July 15, 1997}

\maketitle

\begin{abstract}

We use numerical simulations of galaxy clusters in different
cosmologies to study their ability to form large arcs. The
cosmological models are: Standard CDM (SCDM; $\Omega_0=1$,
$\Omega_\Lambda=0$); $\tau$CDM with reduced small-scale power
(parameters as SCDM, but with a smaller shape parameter of the power
spectrum); open CDM (OCDM; $\Omega_0=0.3$, $\Omega_\Lambda=0$); and
spatially flat, low-density CDM ($\Lambda$CDM; $\Omega_0=0.3$,
$\Omega_\Lambda=0.7$). All models are normalised to the local number
density of rich clusters. Simulating gravitational lensing by these
clusters, we compute optical depths for the formation of large
arcs. For large arcs with length-to-width ratio $\ge10$, the optical
depth is largest for OCDM. Relative to OCDM, the optical depth is
lower by about an order of magnitude for $\Lambda$CDM, and by about
two orders of magnitude for S/$\tau$CDM. These differences originate
from the different epochs of cluster formation across the cosmological
models, and from the non-linearity of the strong lensing effect. We
conclude that only the OCDM model can reproduce the observed arc
abundance well, while the other models fail to do so by orders of
magnitude.


\end{abstract}

\section{Introduction}

Galaxy clusters form differently in different cosmological
models. Their formation history and their internal structure are
influenced by the cosmological parameters. In dense model universes,
$\Omega_0\la1$, clusters form at significantly lower redshifts than in
low-density model universes (e.g.\ Richstone, Loeb, \& Turner 1992;
Bartelmann, Ehlers, \& Schneider 1993; Lacey \& Cole 1993, 1994). The
cosmological constant has a fairly moderate influence on the formation
timescale. Delayed formation is reflected in the abundance of cluster
substructure. Moreover, central cluster densities are higher in
clusters that form earlier.

It is currently unclear whether the different degrees of cluster
substructure expected in cosmological models with different mean
densities lead to observational consequences that can significantly
distinguish between high- and low-density universes. While earlier
studies found cluster X-ray morphologies and density profiles to
differ significantly between different cosmologies (Evrard et al.\
1993; Crone, Evrard, \& Richstone 1994; Mohr et al.\ 1995; Crone,
Evrard, \& Richstone 1996), more recent work concluded that X-ray
morphologies of clusters at the present epoch are fairly similar in
different cosmological models, rendering significant distinctions
difficult (Jing et al.\ 1995). The issue of constraining cluster
shapes using their X-ray emission was also addressed in detail by
Buote \& Tsai (1995a,b). The weak gravitational lensing effect allows
a measurement of the morphology of the projected mass (Wilson, Cole,
\& Frenk 1996; Schneider \& Bartelmann 1997) and constrains cluster
density profiles (Crone et al.\ 1997), but it remains to be shown with
realistic numerical cluster models whether this method provides a more
sensitive tool to quantify cluster morphologies than that provided by
X-ray emission.

An alternative tool is offered by the strong gravitational lens
effect. In order to be strong lenses (i.e.\ in order to produce
appreciable numbers of large arcs from background sources), clusters
have to satisfy several criteria. First, they need to be compact, that
is, their central surface mass densities need to surmount the critical
surface mass density for lensing. The latter depends on redshift. For
background sources at redshifts $z_{\rm s}\sim1$, clusters at
redshifts $0.2\la z_{\rm c}\la0.4$ are the most efficient lenses. If
arcs are to be produced in abundance, a sufficiently large number of
concentrated clusters must be in place at those redshifts. Second,
strong lensing is a highly non-linear effect. This is mainly because
the number of strong lensing events depends sensitively on the number
of cusps in, and the length of, the caustic curves of the
lenses. Cusps require asymmetric lenses. Asymmetric, substructured
clusters are thus much more efficient in producing large arcs than
symmetric clusters, provided the individual cluster sublumps are
compact enough (Bartelmann, Steinmetz, \& Weiss 1995). Both arguments
show that the influence of cosmology on the structure and the
formation timescale of clusters should strongly affect their ability
to form large arcs. Clusters that are being assembled from compact
subclusters at redshifts where lenses are most efficient, $0.2\la
z\la0.4$, should produce many more arcs than clusters which form at
later redshifts from sublumps which are less compact.

Does this line of reasoning imply that different cosmological models
can be distinguished by the number of arcs that are expected in them?
More precisely, does the higher compactness of clusters in low-density
universes, and the later formation time of clusters in high-density
universes, lead to such different numbers of large arcs that limits on
cosmological parameters could be obtained from counting arcs? This is
the question addressed by this paper.

In order to pursue it, we use galaxy clusters simulated in a variety
of cosmological models. The simulations are described in
Sect.~\ref{sec:2}. The simulated clusters are then investigated as to
their strong-lensing effects, following the prescription in
Sect.~\ref{sec:3}. Results are presented in Sect.~\ref{sec:4}, and
summarised in Sect.~\ref{sec:5}. The paper concludes with a discussion
in Sect.~\ref{sec:6}.

Wu \& Mao (1996) already considered the influence of $\Lambda$ on arc
statistics, however with spherically symmetric, non-evolving
clusters. They found that this cluster model predicts $\sim2$ times
more arcs in a low-density, spatially flat universe ($\Omega_0=0.3$,
$\Omega_\Lambda=0.7$) than in an Einstein-de Sitter universe, and that
the latter model falls short by a factor of $\sim4$ to explain the
observed arc abundance. Using a similar model, Hamana \& Futamase
(1997) pointed out that the expected number of observed arcs increases
when the evolution of the background sources is taken into
account. Hattori, Watanabe, \& Yamashita (1997) included observational
selection effects in a study of arc statistics using spherically
symmetric cluster models. Using CDM cluster models at a single
redshift in an Einstein-de Sitter universe, van Kampen (1996) studied
the influence of the normalisation of the power spectrum on the number
of arcs per cluster and concluded that the normalisation should be
somewhat higher than derived from cluster abundance.

\section{Cluster simulations}
\label{sec:2}

We use two different sets of cluster simulations. A first set of five
clusters is taken from high-resolution cosmological simulations kindly
made available by the GIF collaboration, and a second set of four
clusters was simulated for the purpose of studying cluster mass
profiles in different cosmologies. Different numerical techniques were
used for the two simulation sets. They have in common that within each
set, the same random phases are used for the initial density field in
all cosmological models studied, so that the clusters can be compared
individually. The normalisation of the power spectra agrees
approximately with the normalisation to the local cluster number
density. Apart from improved statistics, the approach using two
differently simulated cluster sets allows us to test whether different
numerical techniques yield different results for arc statistics.
Although the normalisations for the two simulation sets are slightly
different, the arc statistics obtained from the two sets individually
agree well with each other. The clusters show density profiles
$\rho(r)$ that are well fitted by the two-parameter function suggested
by Navarro, Frenk, \& White (1996),
\begin{equation}
  \rho(r) = \frac{\rho_{\rm s}}{x\,(1+x)^2}\;,\quad
  x = \frac{r}{r_{\rm s}}\;.
\label{eq:1.1}
\end{equation}
Navarro et al.\ (1996) introduced the concentration parameter
$c=r_{200}/r_{\rm s}$, where $r_{200}$ is the radius enclosing an
average overdensity of 200 times the cosmic background density. At
$z=0$, we find
\begin{equation}
  c \approx \left\{\begin{array}{ll}
    5 & \hbox{for S/$\tau$CDM} \\
    7 & \hbox{for $\Lambda$CDM} \\
    9 & \hbox{for OCDM} \\
  \end{array}\right.\;.
\label{eq:1.2}
\end{equation}
The different values for $c$ reflect the different cluster formation
times. When clusters form earlier, their concentration is higher.

\subsection{First cluster sample: GIF simulations}

\subsubsection{The GIF project}

The GIF project is a joint effort of astrophysicists from Germany and
Israel. Its primary goal is to study the formation and evolution of
galaxies in a cosmological context using semi-analytical galaxy
formation models embedded in large high-resolution $N$-body
simulations. This is done by constructing merger trees of particle
haloes from dark-matter only simulations and placing galaxies into
them using a phenomenological modelling (for a detailed description of
this procedure as well as results cf.\ Kauffmann et al.\ 1997). In
order to achieve both good statistics and an accurate treatment of
early epochs, high resolution simulations are needed which
nevertheless contain a fair sample of the Universe, thus accounting
correctly for the influence of large-scale structure on galaxy
formation. Those characteristics also make these simulations suitable
for the present project.


\subsubsection{The simulations}

The code used for the GIF simulations is called Hydra. It is a
parallel adaptive particle-particle particle-mesh (AP$^3$M) code (for
details on the code cf.\ Couchman, Thomas, \& Pearce 1995; Pearce \&
Couchman 1997). The current version was developed as part of the VIRGO
supercomputing project and was kindly made available by them for the
GIF project. The simulations were started on the CRAY T3D at the
Computer Centre of the Max-Planck Society in Garching (RZG) on 128
processors. Once the clustering strength required an even larger
amount of total memory, they were transferred to the T3D at the
Edinburgh Parallel Computer Centre (EPCC) and finished on 256
processors.

A set of four simulations with $N=256^3$ and with different
cosmological parameters was run. Apart from the fiducial Cold Dark
Matter (CDM) scenario, denoted SCDM, which has $\Omega_0=1$ and
$h=0.5$\footnote{As usual, the Hubble constant is written as
$H_0=100\,h\;$ ${\rm km\,s^{-1}\,Mpc^{-1}}$.}, another $\Omega_0=1$
and $h=0.5$ model was run ($\tau$CDM) which has the same shape
parameter for the power spectrum, $\Gamma=0.21$, as the remaining two
models. Those are both models with $\Omega_0=0.3$, the first one being
a flat model with a cosmological constant ($\Lambda$CDM,
$\Omega_\Lambda=0.7$, $h=0.7$), and the last model being an open model
(OCDM, $\Omega_\Lambda=0$, $h=0.7$). The case of the $\tau$CDM model
is particularly interesting because it shares the $\Omega_0=1$
dynamics with the SCDM model, but has a power spectrum with the same
shape parameter as the two low-$\Omega_0$ models. A value of
$\Gamma=0.21$ is usually preferred by analyses of galaxy clustering,
cf.\ Peacock \& Dodds (1994). This is achieved in the $\tau$CDM model
despite $\Omega_0=1$ and $h=0.5$ by assuming that a massive neutrino
(usually taken to be the $\tau$ neutrino) had existed during the very
early evolution of the Universe. It must have decayed later, thus
shifting the epoch when matter started to dominate over radiation in
the Universe, and the neutrino mass and lifetime are chosen such that
$\Gamma=0.21$. For a detailed description of such a model see White,
Gelmini, \& Silk (1995).

The GIF simulations adopt the power spectrum
\begin{equation}
  P(k) = \frac{Ak}{[1+[ak +(bk)^{3/2} + (ck)^2]^{\nu}]^{2/\nu}}\,
\label{eq:2.1}
\end{equation}
with
\begin{eqnarray}
  a &=& 6.4\,\Gamma^{-1}\,h^{-1}\,{\rm Mpc}\;,\nonumber\\
  b &=& 3.0\,\Gamma^{-1}\,h^{-1}\,{\rm Mpc}\;,\nonumber\\
  c &=& 1.7\,\Gamma^{-1}\,h^{-1}\,{\rm Mpc}\;,\nonumber\\
 \nu &=& 1.13\nonumber\\
\label{eq:2.2}
\end{eqnarray}
(Bond \& Efstathiou 1984). In order to completely fix $P(k)$, the
normalisation, $A$, has to be chosen. This can be done on the basis of
measurements of the microwave background anisotropies by the COBE
satellite. However, this approach suffers from the fact that COBE
measured fluctuations on scales much larger than those pertinent to
the simulations. Hence, one has to assume that there is no additional
physics that could alter the result like, e.g., gravitational waves or
a slight tilt of the initial spectrum away from the scale-invariant
form. The approach taken here avoids this problem. The mass function
of objects in the Universe is very steep at the high-mass end. In
other words, massive objects (like clusters of galaxies) are not only
rare, but their {\em abundance\/} sensitively depends on the amplitude
of the power spectrum. White, Efstathiou, \& Frenk (1993) introduced
this way of fixing the amplitude by determining $\sigma_8$, the square
root of the variance of the density field smoothed over
$8\,h^{-1}$\,Mpc spheres, such that the observed abundance of rich
clusters is matched. They used the cluster mass function. Recent
studies of the cluster X-ray temperature function (Eke, Cole, \& Frenk
1996; Viana \& Liddle 1996) find similar results. For the low-density
GIF simulations, the result by Eke et al.\ (1996) was taken,
\begin{equation}
  \sigma_8 = \left\{\begin{array}{ll}
    (0.52\pm 0.04)\,\Omega_0^{-0.46+0.10\,\Omega_0} &
    \mbox{for $\Omega_\Lambda=0$} \\
    (0.52\pm 0.04)\,\Omega_0^{-0.52+0.13\,\Omega_0} &
    \mbox{for $\Omega_0+\Omega_\Lambda=1$}
  \end{array}\right.
\label{eq:2.3}
\end{equation}
For the $\Omega_0=1$ simulations, slightly larger values than
suggested by eq.~(\ref{eq:2.3}) were adopted, according to the earlier
result by White et al.\ (1993). Table \ref{tab:2.1} summarises the
model parameters.

\begin{table}
\caption{Cosmological parameters of the GIF models. $\Omega_0$ and
  $\Omega_\Lambda$ are the density parameters for matter and the
  cosmological constant, $h$ is the Hubble parameter, $\sigma_8$ is
  the variance of the density field in spheres of $8\,h^{-1}\;$Mpc,
  and $\Gamma$ is the shape parameter of the power spectrum. Also
  given are the size of the cosmological simulation box and the mass
  $M_{\rm max}$ within $1.5\,h^{-1}\;{\rm Mpc}$ radius of the most
  massive cluster in $10^{15}\,M_\odot$.}
\label{tab:2.1}
\medskip
\centering
\begin{tabular}{l*{7}{c}}
\hline
Model & $\Omega_0$ & $\Omega_\Lambda$ & $h$ & $\sigma_8$ &
$\Gamma$ & Box Size & $M_{\rm max}$ \\
& & & & & & [${\rm Mpc}/h$] & \\
\hline
SCDM1         & 1.0 & 0.0 & 0.5 & 0.60 & 0.50 &  85 & 0.74 \\
$\tau$CDM1    & 1.0 & 0.0 & 0.5 & 0.60 & 0.21 &  85 & 0.74 \\
$\Lambda$CDM1 & 0.3 & 0.7 & 0.7 & 0.90 & 0.21 & 141 & 0.84 \\
OCDM1         & 0.3 & 0.0 & 0.7 & 0.85 & 0.21 & 141 & 0.85 \\
\hline
\end{tabular}
\end{table}

The parameters shown in Table~\ref{tab:2.1} were chosen not only to
fulfil cosmological constraints, but also to allow a detailed study of
the clustering properties at very early redshifts. The masses of
individual particles are $1.0\times10^{10}\,h^{-1}\,M_{\odot}$ and
$1.4\times10^{10}\,h^{-1}\,M_{\odot}$ for the high- and low-$\Omega_0$
models, respectively. The gravitational softening was taken to be
$30\,h^{-1}\,{\rm kpc}$.

Clusters are obtained from the simulation as follows. High-density
regions are searched using a standard friends-of-friends group finder
with a linking length of $b=0.05$ times the mean interparticle
separation. This selects only the dense cores of any collapsed
object. Around the centres of these, all particles are collected which
lie within a sphere of radius $r_{\rm A}=1.5\,h^{-1}\,{\rm Mpc}$,
which corresponds to Abell's radius.  These objects are taken as
clusters. For our analysis, the five most massive clusters are cut out
of the simulation volumes. This procedure needs to be expanded if the
centres of two large clusters are closer together than $r_A$. In this
case usually the more massive cluster is taken, and the other one is
deleted from the list. In our case, however, this problem did not
occur. The initial density fields for the different cosmologies share
the same random phases.

\subsection{Second cluster sample}

The second cluster sample is simulated using a special multi-mass
technique which is explained in detail in Huss, Jain, \& Steinmetz
(1997). In contrast to the GIF simulations, this technique gives only
one massive cluster per run. However, it allows one to study the
evolution of one individual cluster without the need for extensive
computer resources.

The essential part of the multi-mass technique is the initial particle
arrangement. It consists of three spherical layers embedded into a
cubic volume, each filled with particles of different mass. The
central sphere encompasses the least massive particles and is
surrounded by two shells of more massive particles. The rest of the
cubic simulation volume is filled up with the most massive
particles. The inner sphere must initially be large enough to enclose
all particles which end up in a cluster. The gravitational forces on
the particles are calculated using a combined GRAPE/PM $N$-body code
assuming periodic boundary conditions. The PM part performs force
calculations with periodic boundary conditions for all particles. In
the inner three shells, the force is additionally calculated with a PM
code using vacuum boundary conditions. This force is subtracted from
the periodically computed PM force to obtain the periodic contribution
to the force only. This is added to the highly resolved force provided
by the GRAPE board for the particles in the inner shells.

The second cluster sample consists of 12 clusters in total. Four
clusters are simulated for each of three different cosmologies, which
resemble the SCDM, $\Lambda$CDM, and OCDM models of the GIF
project. The model parameter are summarised in
Table~\ref{tab:2.2}. All clusters belonging to one cosmological model
are part of the same realisation of the corresponding density
field. In addition, the phases of the initial Gaussian random field
are identical in the three cosmological models.

\begin{table}
\caption{Cosmological parameters of the models for the second cluster
  sample. The meaning of the symbols is the same as in
  Tab.~\protect\ref{tab:2.1}.}
\label{tab:2.2}
\medskip
\centering
\begin{tabular}{l*{7}{c}}
\hline
Model & $\Omega_0$ & $\Omega_\Lambda$ & $h$ & $\sigma_8$ &
$\Gamma$ & Box size & $M_{\rm max}$ \\
& & & & & & [${\rm Mpc}/h$] & \\
\hline 
SCDM2         & 1.0 & 0.0 & 0.5 & 0.60 & 0.50 & 144   & 0.75 \\
$\Lambda$CDM2 & 0.3 & 0.7 & 0.7 & 1.12 & 0.21 & 201.6 & 1.67 \\
OCDM2         & 0.3 & 0.0 & 0.7 & 1.12 & 0.21 & 201.6 & 1.41 \\
\hline
\end{tabular}
\end{table}

The initial conditions are calculated using eq.~(\ref{eq:2.1}) to
match the power spectrum of the different cosmologies. The
normalisation of $P(k)$ is chosen as determined by White et al.\
(1993) for SCDM2, and for OCDM2 and $\Lambda$CDM2 it matches the COBE
CMB anisotropy measurements. Contributions from gravitational waves
need not be considered since eq.~(\ref{eq:2.1}) assumes that the
primordial power spectrum after inflation is of the
Harrison-Zel'dovich form. The COBE normalisation yields a slightly
higher $\sigma_8$ than that derived by Eke et al.\ (1996). The box
size $L$ of the simulation volume is fixed to $288\;$Mpc in physical
coordinates for each run to achieve the same physical spatial
resolution in all models. The mass resolution in the central sphere is
$4.9\times10^{10}\,h^{-1}\;M_\odot$ for the high-density model, and
$2.9\times10^{10}\,h^{-1}\;M_\odot$ for the low-density models.

Since the simulations result in one massive cluster in the
high-resolution region, it can easily be identified by looking for the
deepest potential well in this region. However, particular attention
has to be paid in order to set up initial conditions with a suitable
overdense region in the central sphere, representing the seed for the
massive cluster. This is done in the following way. First, a
pre-simulation is performed by filling the whole simulation box with
particles of the same mass as those in the second layer. At $z=0$,
the cluster-like objects are identified using a special group-finding
algorithm (Huss et al.\ 1997). The particles finally forming these
objects define the corresponding overdense region at the initial
redshift. The final starting configuration is then centred on one of
these regions. By adding small scale power to such a density peak, the
clustering properties of the region can be changed. Hence, it is
possible that the simulation finally arrives at several low-mass
objects rather than at one massive halo. This can be avoided by
testing the clustering properties using the Zel'dovich
approximation. When propagated to low redshifts with this technique,
the particles in the central sphere must form a distinct matter
accumulation in the centre rather than showing only filamentary
structure.

With this procedure, four suitable overdense regions are identified in
the SCDM2 model. For the $\Lambda$CDM2 and for the OCDM2 clusters, the
starting configurations are centred on the same regions as for the
SCDM2 clusters. This is possible since the clustering properties are
defined mainly by the local realisation of the random field, which is
the same for all three models. However, the final cluster haloes need
not represent the most massive clusters in the simulation box.

\section{Simulations of arcs}
\label{sec:3}

Our method to investigate the arc-formation statistics of the
numerical cluster models was described in detail by Bartelmann \&
Weiss (1994). We will therefore keep the present description brief and
refer the reader to that paper for further information. For general
information on gravitational lensing, see Schneider, Ehlers, \& Falco
(1992) or Narayan \& Bartelmann (1997), and references therein.

The numerical cluster models yield the spatial coordinates and
velocities of discrete particles with equal mass. In order to use them
for gravitational lensing, we need to compute the surface mass density
distribution of each cluster model in each of the three independent
directions of projection. The mass density is first determined by
sorting the particles into a three-dimensional grid and subsequently
smoothing with a Gaussian filter function. The grid resolution and the
width of the Gaussian are adapted to the numerical resolution of the
$N$-body codes in order not to lose spatial resolution by the
smoothing of the density field. The smoothed density field is then
projected onto the three sides of the computation volume to obtain
three surface-density fields for each cluster.

The physical surface mass density fields are then scaled by the
critical surface mass density for lensing, which apart from the
cosmological parameters depends on the cluster- and source
redshifts. We keep the redshift for all sources fixed at $z_{\rm
s}=1$, and the clusters are at $0<z_{\rm c}<1$. This finally yields
three two-dimensional convergence fields $\kappa(\vec x,z_{\rm c})$
for each cluster model at redshift $z_{\rm c}$. From $\kappa(\vec
x,z_{\rm c})$, all quantities determining the local lens mapping,
i.e., the deflection angle and its spatial derivatives, can be
computed. We determine the lens properties of the clusters on grids
with an angular resolution of $0\farcs3$ in the lens plane in order to
ensure that lensed images be properly resolved.

Sources are then distributed on a regular grid in the source
plane. The resolution of this source grid can be kept low close to the
field boundaries because there no large arcs occur. Close to the
caustics of the clusters, where the large arcs are formed, the
source-grid resolution is increased with the increasing strength of
the lens. For our later purpose of statistics, sources are weighted
with the inverse resolution of the grid on which they are placed. The
sources are taken to be intrinsically randomly oriented ellipses with
their axis ratios drawn randomly from the interval $[0.5,1]$, and
their axes determined such that their area equals that of circles with
radius $0\farcs5$. Although this choice of source properties appears
fairly simple, it should not affect the arc statistics because these
mainly reflect the local properties of the lens mapping, which are
independent of the particular choice of source size or ellipticity
distribution. We checked that a change in average source size did not
change the results.

The sources are then viewed through the cluster lenses. All images are
classified in the way detailed by Bartelmann \& Weiss (1994). Among
other things, the classification yields for each image its length $L$,
its width $W$, and its curvature radius $R$. In total, we classify the
images of about $1.3\times10^6$ sources.

Knowing the area covered by the cluster fields, and having determined
the frequency of occurrence of image properties such as a given
length, width, and curvature radius, we can compute cross sections
$\sigma$ for the formation of images with such properties. The arc
cross section of a cluster is defined as the area in the {\em source
plane\/} within which a source has to lie in order to be imaged as an
arc. We mostly focus on cross sections for the length-to-width ratio
$r$ of arcs. Apart from the image properties, the cross sections
depend on redshift, $\sigma=\sigma(z)$.

Given cross sections $\sigma(z)$, we compute the optical depth $\tau$
for the formation of large arcs. The optical depth is the fraction of
the entire source plane which is covered by cluster cross sections,
\begin{equation}
  \tau = \frac{n_{\rm c}}{4\pi D_{\rm s}^2}\,
  \int_0^{z_{\rm s}}\,{\rm d}z\,(1+z)^3\,
  \left|\frac{{\rm d}V(z)}{{\rm d}z}\right|\,\sigma(z)\;,
\label{eq:3.1}
\end{equation}
where $n_{\rm c}$ is the present cluster number density, $D_{\rm s}$
is the angular-diameter distance to the source plane, and ${\rm
d}V(z)$ is the proper volume of a spherical shell of width ${\rm d}z$
about $z$. The factor $(1+z)^3$ accounts for the cosmological
expansion factor.

\section{Results}
\label{sec:4}

\subsection{Cross sections}

\begin{figure}[ht]
  \centering\mbox{\epsfxsize=\fsize\epsffile{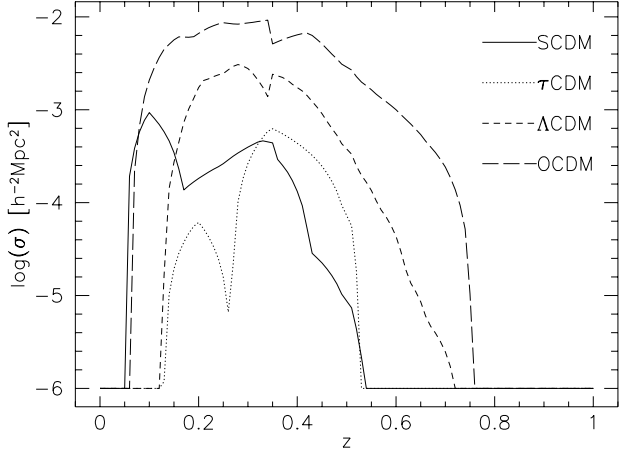}}
\caption{Averaged cross sections for arc length-to-width ratio
  $r\ge7.5$ for clusters in the four different cosmological models,
  distinguished by line types as indicated. The figure shows that
  clusters in the $\tau$CDM model produce the fewest arcs, and
  clusters in the open CDM model the most. Note the logarithmic scale
  of the ordinate: The maxima of the cross sections differ by more
  than an order of magnitude. The redshift ranges where $\sigma(z)>0$
  are larger in O/$\Lambda$CDM than in S/$\tau$CDM.}
\label{fig:4.1}
\end{figure}

According to the prescription in Sect.~\ref{sec:3}, we first calculate
cross sections for each individual model cluster, projected in each of
the three independent spatial directions. Interpolating in redshift
between the redshifts of the model clusters, this yields cross
sections $\sigma(z)$ as a function of redshift. We then average these
cross sections (1) over the three projection directions and (2) over
all model clusters within a given cosmological model. We thus obtain
cross sections for the four cosmological models. Figure \ref{fig:4.1}
shows an example, the cross sections for arcs with length-to-width
ratio $r\ge7.5$.

The averaged cross sections in Fig.~\ref{fig:4.1} reveal huge
differences between the cosmological models. While the cross sections
for standard CDM (SCDM) and $\tau$CDM are comparable, the maximum
cross sections for $\Lambda$CDM and open CDM (OCDM) exceed that for
SCDM by about half and one order of magnitude, respectively, and the
redshift range where $\sigma(z)>0$ is wider in O/$\Lambda$CDM than in
S/$\tau$CDM. Cross sections for other, large values of the
length-to-width ratio $r$, or for large arc lengths, show a
qualitatively similar behaviour.

Since the clusters in different cosmologies arise from initial density
perturbations with the same random phases, they can also be compared
individually rather than statistically. On the whole, the individual
clusters show the same qualitative behaviour as the averaged cross
sections shown in Fig.~\ref{fig:4.1}. Results obtained for each
cluster set individually agree well with each other.

\subsection{Arc-cluster redshift}

\begin{figure}[ht]
  \centering\mbox{\epsfxsize=\fsize\epsffile{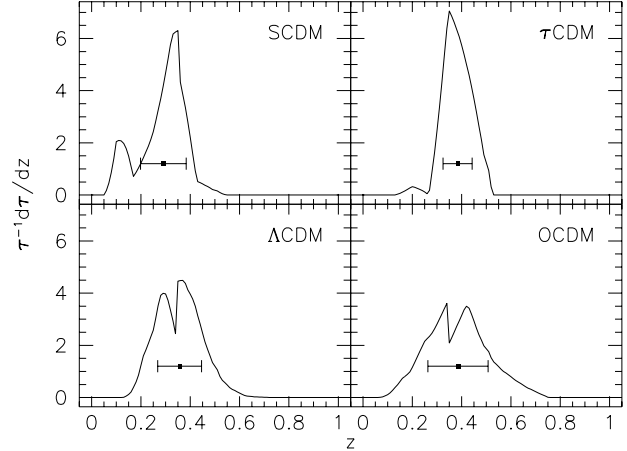}}
\caption{Normalised, differential optical depth as a function of
  redshift, for arcs with length-to-width ratio $r\ge7.5$. The curves
  indicate the most probable redshift for a cluster forming arcs for
  the four cosmological models used. The bars show the 1-$\sigma$
  redshift range, the dots indicate the mean arc-cluster redshift. The
  plot shows that there is no significant difference in arc cluster
  redshift between the four cosmological models.}
\label{fig:4.2}
\end{figure}

At what redshifts do we expect to find the most clusters that produce
large arcs? In other words, clusters at which redshift contribute most
to the arc optical depth? To answer this question, we compute the
optical depth $\tau$ from eq.~(\ref{eq:3.1}) and the differential
optical depth ${\rm d}\tau/{\rm d}z$, and plot in Fig.~\ref{fig:4.2}
the normalised differential optical depth $\tau^{-1}\,{\rm d}\tau/{\rm
d}z$ as a function of redshift for the four cosmological models.

The curves in Fig.~\ref{fig:4.2} show that the differential optical
depth peaks around $z\sim0.3-0.4$. The bars inserted in the figure
indicate the 1-$\sigma$ redshift range, and the dots show the average
arc-cluster redshift. Although there is a slight tendency that the
mean arc-cluster redshift is smallest in the SCDM model, larger for
$\Lambda$CDM, and largest for $\tau$CDM and OCDM, the redshift
variances are large enough for the redshift ranges in the cosmological
models to overlap. The figure furthermore suggests that the
differences between cosmological models are dominated by
noise. Numbers are given in Table~\ref{tab:4.1}.

\begin{table}
\caption{Mean redshifts and redshift ranges for clusters producing
  large arcs in the four different cosmological models.}
\label{tab:4.1}
\medskip
\centering
  \begin{tabular}{ccc}
  \hline
  model & $\bar{z}_{\rm c}$ &
  $\langle(z_{\rm c}-\bar{z}_{\rm c})^2\rangle^{1/2}$ \\
  \hline
  SCDM &         $0.29$ & $0.09$ \\
  $\tau$CDM &    $0.38$ & $0.06$ \\
  $\Lambda$CDM & $0.36$ & $0.09$ \\
  OCDM &         $0.39$ & $0.12$ \\
  \hline
  \end{tabular}
\end{table}

\subsection{Optical depth}

\begin{figure}[ht]
  \centering\mbox{\epsfxsize=\fsize\epsffile{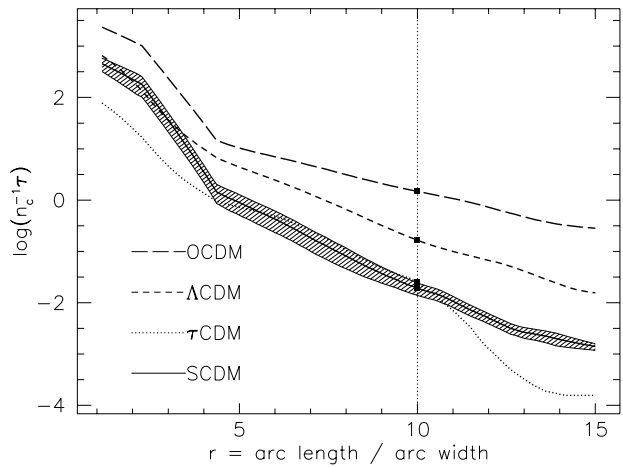}}
\caption{Optical depth for arc formation as a function of arc
  length-to-width ratio $r$, for the four different cosmological
  models. The optical depth for the $\tau$CDM and the standard CDM
  models are comparable for intermediate length-to-width ratios
  $r$. For large $r$, the optical depth is smallest for the $\tau$CDM
  model. The largest optical depth is produced by clusters in the open
  CDM model, followed by those in the $\Lambda$CDM model. For
  $r\sim10$, the optical depths for open CDM, $\Lambda$CDM, and
  standard CDM differ by about an order of magnitude each. The hatched
  area around the SCDM curve indicates 1-$\sigma$ bootstrap errors.}
\label{fig:4.3}
\end{figure}

We now compare the optical depth $\tau$ for formation of arcs with
given length-to-width ratio $r$ in the four cosmological models. As
before, the optical depth is calculated from eq.~(\ref{eq:3.1}). We
do not specify the cluster number density $n_{\rm c}$ yet, but
calculate the optical depth per unit cluster density, $n_{\rm
c}^{-1}\tau$. Results are shown in Fig.~\ref{fig:4.3}.

Figure \ref{fig:4.3} confirms the trends indicated by the cross
sections in Fig.~\ref{fig:4.1}, but allows to compare optical depths
for a wide range of arc length-to-width ratios $r$. There is an
interval at intermediate $r$, $5\la r\la10$, where the optical depths
for SCDM and $\tau$CDM are almost equal. Only at $r\ga10$ does the
optical depth in $\tau$CDM models drop below that of SCDM models. For
$r\ga4$, the optical depth for $\Lambda$CDM models is constantly
higher than that for SCDM models by a factor of $\sim10$. The optical
depth for the OCDM model is highest, exceeding the SCDM value by up to
$\sim2$ orders of magnitude at large $r$.

The hatched region around the SCDM curve illustrates 1-$\sigma$
bootstrap errors, which we obtained by bootstrapping the cluster
sample. They give an impression of the uncertainty of the optical
depth due to the limited number of clusters in our samples. The
uncertainty due to the specific realisation of the lensed background
galaxy sample are smaller by about a factor of five.

\begin{figure}[ht]
  \centering\mbox{\epsfxsize=\fsize\epsffile{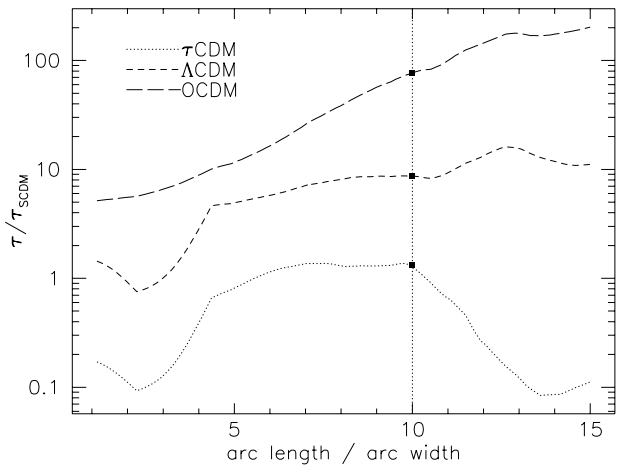}}
\caption{Optical depths for large arcs, normalised by the arc optical
  depth in the standard CDM model. The curves emphasise the large
  difference in optical depth for the four models. For the largest $r$
  plotted, there are more than two orders of magnitude between open
  CDM and standard CDM clusters, about one order of magnitude between
  $\Lambda$CDM and standard CDM, and clusters in the $\tau$CDM model
  are less efficient than the standard CDM models for
  $r\protect\ga10$.}
\label{fig:4.4}
\end{figure}

In order to emphasise the results, Fig.~\ref{fig:4.4} shows the
optical depths for $\tau$CDM, $\Lambda$CDM, and OCDM, divided by the
optical depth for SCDM.

\section{Discussion}
\label{sec:5}

\subsection{Results}

We have used numerical simulations to calculate the optical depth for
the formation of large arcs in different cosmologies. The simulated
clusters are cut out of large cosmological simulation volumes, so that
the tidal effects of surrounding matter are taken into account. The
cosmological simulations were normalised to the observed local number
density of rich galaxy clusters. Four different cosmogonic models were
used. These are: standard CDM (SCDM), with $\Omega_0=1$,
$\Omega_\Lambda=0$, $h=0.5$, $\sigma_8=0.6$, and shape parameter
$\Gamma=0.5$; a CDM model with reduced small-scale power ($\tau$CDM),
which differs from SCDM only by the shape parameter $\Gamma=0.21$;
open CDM (OCDM) with $\Omega=0.3$, $\Omega_\Lambda=0$, $h=0.7$,
$\sigma_8=0.9$ or $1.1$, and $\Gamma=0.21$; and finally a spatially
flat, low-density CDM model ($\Lambda$CDM) with $\Omega_0=0.3$,
$\Omega_\Lambda=0.7$, $h=0.7$, $\sigma_8=0.9$ or $1.1$, and
$\Gamma=0.21$. All cosmological simulations start from density
perturbations with the same random phases, so that all clusters can be
compared individually in different cosmogonies. For SCDM, OCDM, and
$\Lambda$CDM, we simulated nine clusters, and five clusters for
$\tau$CDM.

The lensing properties of the clusters with respect to large arcs were
calculated in a way that has extensively been described earlier
(Bartelmann \& Weiss 1994; Bartelmann et al.\ 1995). The calculations
result in averaged cross sections for the clusters as a function of
redshift, which can then be converted to optical depths $\tau$ for the
formation of large arcs.

Our main result is that the optical depths for large arcs, with
length-to-width ratio $r\ge10$, differ by orders of magnitude for the
different cosmologies. Generally, clusters in the SCDM and $\tau$CDM
models produce the smallest optical depth. For $r\sim10$, the optical
depths for these two models are comparable, but for larger $r$, the
optical depth in the $\tau$CDM model falls below that in the SCDM
model. In the $\Lambda$CDM model, the arc optical depth is larger by
about an order of magnitude than for SCDM, and the optical depth is
largest in the OCDM model, exceeding the SCDM optical depth by about
two orders of magnitude. We emphasise that these results are
independent of whether our cluster samples are in any sense {\em
complete\/} or not, because the simulations are designed such that the
clusters can be compared individually. Our conclusions do, however,
rest on the assumption that the simulated clusters are {\em typical\/}
for the clusters with the largest mass in each of the cosmological
models. We believe that this is guaranteed by the large size of the
cosmological simulation volumes from which the clusters were taken.

It is a combination of effects that leads to the large difference in
arc optical depth across the cosmological models that we have
investigated. (i) Clusters form earlier in low-density than in
high-density universes. In SCDM, normalised to the cluster abundance,
the formation of such clusters which would in principle be massive
enough for strong lensing is delayed to such low redshifts that they
fail to be efficient lenses for sources at redshifts $z_{\rm
s}\sim1$. (ii) For low-density universes, the proper volume per unit
redshift is larger than for high-density universes. Given the observed
number density of clusters today, this volume effect increases the
number of clusters between the source sphere and the observer when
$\Omega_0$ is small. (iii) Clusters that form early are more
concentrated than clusters that form late. Clusters in low-density
universes therefore reach higher central surface mass densities than
clusters in high-density universes. (iv) Strong gravitational lensing
is a highly non-linear effect. This is because the arc cross section
of a cluster sensitively depends on the length of the caustic curve
and the number of cusp points contained in it. The properties of the
caustic curve do not only depend on the surface mass density, but also
on the tidal field of a cluster, which is influenced by the cluster
morphology. (v) Because of (iv), asymmetric clusters are much more
efficient in producing large arcs than symmetric clusters (Bartelmann
et al.\ 1995). The degree of substructure of a cluster is therefore
very important for arc statistics. While clusters are in the process
of formation, they are expected to be highly asymmetric. If this
happens at redshifts where lensing is efficient for a given source
population, the asymmetric cluster morphology further increases the
strong-lensing cross section. Most clusters in OCDM and $\Lambda$CDM
form at $z\sim0.3$, exactly where lensing is most efficient for
sources at $z_{\rm s}\sim1$. Clusters in SCDM and $\tau$CDM form
later, at $z\sim0.1$, where their lensing efficiency and that of their
sublumps is already suppressed by the lensing geometry. Although
clusters in S/$\tau$CDM form later than in O/$\Lambda$CDM and should
therefore be more asymmetric, the sublumps in O/$\Lambda$CDM clusters
are more compact and thus tend to persist for a longer time after
merging with the cluster.

\subsection{Illustration}

\begin{figure}[ht]
  \centering\mbox{\epsfxsize=\fsize\epsffile{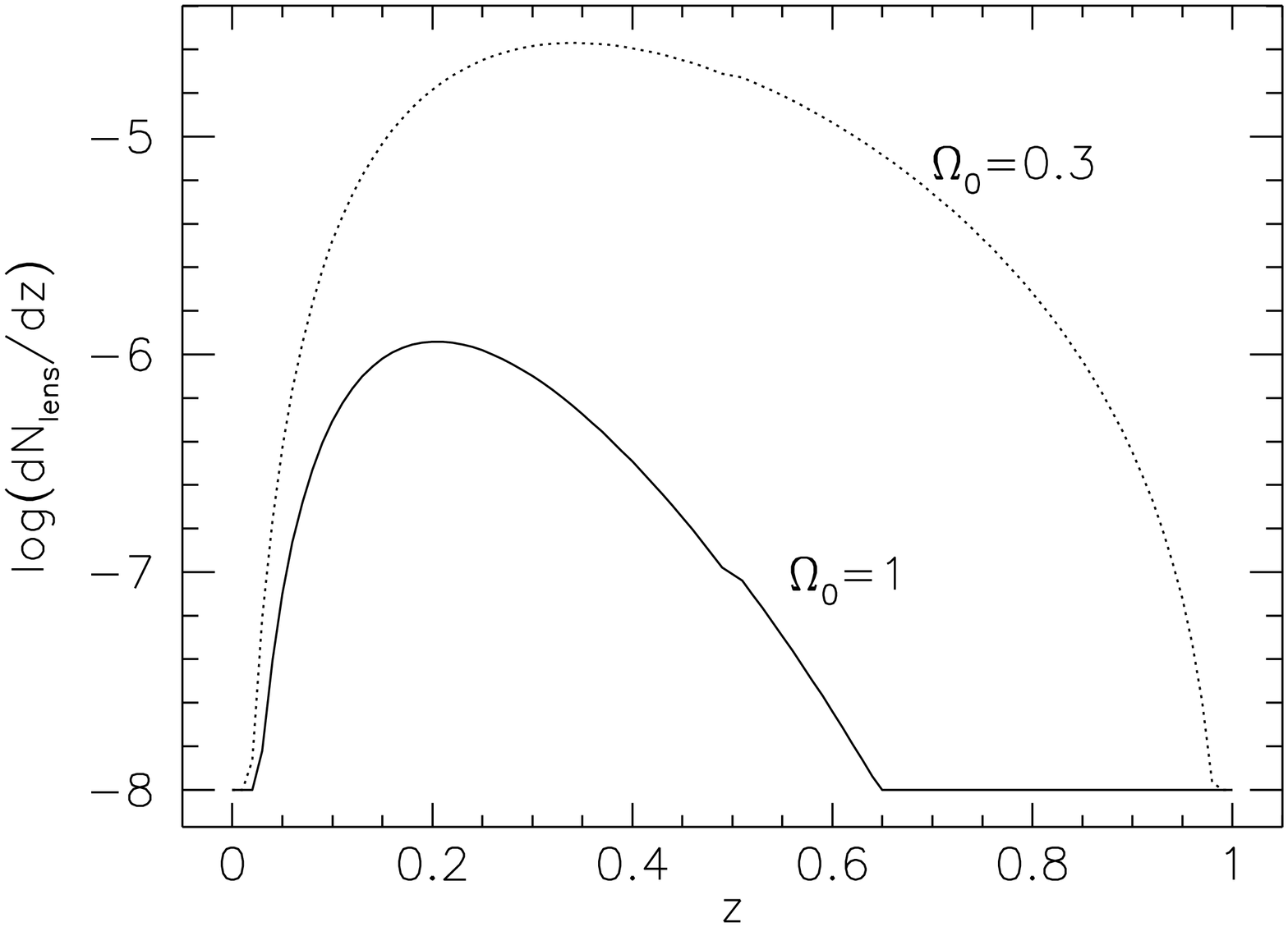}}
\caption{Estimate for the number of efficient lensing clusters per
  redshift interval, ${\rm d}N_{\rm lens}/{\rm d}z$, as given in
  eq.~(\protect\ref{eq:5.2}). Results for $\Omega_0=1$ and
  $\Omega_0=0.3$, both with $\Omega_\Lambda=0$, are plotted, as
  indicated. The figure illustrates that the delayed formation of
  clusters in a high-density universe, combined with the effects of
  lensing efficiency and cosmic volume, already account for a large
  difference in the expected number of arcs.}
\label{fig:5.1}
\end{figure}

A simple Press-Schechter type argument illustrates the influence of
formation time and cosmic volume. According to Press \& Schechter
(1974), the (comoving) fraction of the cosmic matter that is contained
in clusters is
\begin{equation}
  F_{\rm c}(z) = \frac{1}{2}\,{\rm erfc}\left(
  \frac{\delta_{\rm c}}{\sqrt{2}\,\sigma_R\,{\cal D}(z)}
  \right)\;,
\label{eq:5.1}
\end{equation}
with $\delta_{\rm c}\approx1.686$, $\sigma_R$ the variance of the
density contrast on cluster scales today, and ${\cal D}(z)$ the
(cosmology-dependent) linear growth factor of density
perturbations. The cluster fraction at redshift $z$, normalised to the
present cluster fraction, provides an estimate for the change in
cluster number density with redshift. Multiplying with the proper
cosmic volume $4\pi\,D^2(z)\,|{\rm d}(ct)/{\rm d}z|\,{\rm d}z$ of a
shell of width ${\rm d}z$ and the squared effective lensing distance
$D_{\rm eff}(z,z_{\rm s})$ yields an estimate for the number of
efficient lensing clusters per redshift interval,
\begin{equation}
  \frac{{\rm d}N_{\rm lens}}{{\rm d}z} =
  F_{\rm c}(z)\times(1+z)^3\times
  D_{\rm eff}^2(z,z_{\rm s})\times
  4\pi\,D^2(z)\,
  \left|\frac{{\rm d}(ct)}{{\rm d}z}\right|\;,
\label{eq:5.2}
\end{equation}
because the cross section per cluster should scale approximately with
$D^2_{\rm eff}(z,z_{\rm s})$. This quantity is plotted in
Fig.~\ref{fig:5.1} for $\Omega_0=1$ and $\Omega_0=0.3$, with
$\Omega_\Lambda=0$. Of course, this simple estimate completely
neglects the influence of the change in cluster concentration across
the cosmological models, and the non-linearities of the strong lensing
effect. However, it suffices to demonstrate that large differences in
the arc cross section are expected between high- and low-density
universes.

\subsection{Influence of ``missing clusters''}

All cosmological simulation volumes from which we have taken the
cluster models are of order a few times $10^6\,h^{-3}\,{\rm
Mpc}^3$. Since the cluster mass function is very steep, we are
therefore likely to miss the most massive clusters. In order to
estimate their influence on the arc cross sections, we have repeated
the arc simulations for SCDM with surface-mass densities rescaled to
higher cluster mass. Let $M_0$ and $M>M_0$ be the original and
rescaled cluster masses, respectively. Then, we take
\begin{equation}
  \kappa'(\vec x,z_{\rm c},M) =
  \left(\frac{M}{M_0}\right)^{1/3}\,
  \kappa\left[
    \left(\frac{M_0}{M}\right)^{1/3}\vec x,z_{\rm c},M_0
  \right]
\label{eq:5.3}
\end{equation}
for the rescaled surface-mass density and compute the arc cross
section from that. In effect, we calculate the arc cross section
$\sigma(z_{\rm c},M)$ of a cluster of similar structure as the
original one, but with higher total mass. For each cluster model at
redshift $z_{\rm c}$, we then average the arc cross sections over
mass, weighted by the Press-Schechter cluster mass function $n(M)$,
\begin{equation}
  \langle\sigma(z_{\rm c})\rangle = \frac
  {\int{\rm d}M\,n(M)\,\sigma(z_{\rm c},M)}
  {\int{\rm d}M\,n(M)}\;.
\label{eq:5.4}
\end{equation}
Finally, we compute mass-averaged optical depths $\langle\tau\rangle$
by substituting $\langle\sigma(z_{\rm c})\rangle$ for $\sigma(z_{\rm
c})$ in eq.~(\ref{eq:3.1}).

We found that $\langle\tau\rangle$ differs from $\tau$ only by 10-15
per cent. Although the arc cross section is a fairly steep function of
mass, the increase in $\sigma$ is more than compensated by the steep
decrease of the cluster mass function. For example, quadrupling the
masses of the $\tau$CDM clusters increases the averaged cross section
for arcs with $r\ge10$ by about two orders of magnitude, but decreases
the cluster mass function by about three orders of magnitude, almost
completely cancelling the effect of the larger arc cross section. For
clusters in the other cosmological models, the change in the optical
depth should be smaller than that in $\tau$CDM. Clusters with higher
masses than those contained in our sample can therefore safely be
neglected.

We have only studied the most massive clusters found in each
cosmological simulation. Even then, the least massive clusters in each
sample contribute little or nothing to the arc optical
depth. Extending our samples by including less massive clusters would
therefore change the arc optical depth negligibly or not at all.

Since we select the most massive clusters at redshift zero and study
their progenitors, it could be that at higher redshift other clusters
in the cosmological simulations would be more massive. In other words,
it is not immediately clear that the progenitors of the most massive
clusters are also the most massive clusters present at higher
redshift. In order to test that, we selected the $\tau$CDM model,
where clusters form at the lowest redshifts, and checked whether the
progenitors of the our sample of the five most massive clusters is
identical with the sample of the five most massive clusters at the
redshifts relevant for lensing. This turned out to be the case. In the
other models, where clusters form at higher redshifts, possible
misidentifications are even less likely than in $\tau$CDM.

Given these results, we are confident that our cluster samples fairly
reflect those clusters that dominate the optical depth for the
formation of large arcs.

\section{Comparison with observations}
\label{sec:6}

So can we constrain cosmological parameters through arc statistics?
For that, we would have to compare the number of observed arcs to that
predicted by our models. This comparison is hampered by the fact that
there is no complete sample of observed clusters selected by {\em
mass\/}, as it should be for a fair comparison. There is one cluster
sample, however, whose definition comes close to this criterion,
namely the EMSS sample of X-ray bright clusters, for which the X-ray
luminosity in the EMSS energy band is $L_{\rm
X}\ge2\times10^{44}\,{\rm erg\,s^{-1}}$ ($h=0.5$, $q_0=0.5$). The
number density of such clusters is estimated to be $n_{\rm
c}\sim2\times10^{-6}\,h^3\,{\rm Mpc}^{-3}$ (Le F\`evre et al.\
1994). Arc surveys in this sample have shown that the number of arcs
with $r\ge10$ and a limiting magnitude of $B=22.5$ (or $R=21.5$; these
are the arc criteria set up by Wu \& Hammer 1993) is roughly
$\sim0.2-0.3$ per cluster (Le F\`evre et al.\ 1994; Gioia \& Luppino
1994).

Clusters with $L_{\rm X}\ge2\times10^{44}\,{\rm erg\,s^{-1}}$ should
be fairly represented by the massive simulated clusters in our
samples. Having velocity dispersions $\ga800\,{\rm km\,s}^{-1}$, the
empirical relation between velocity dispersion and X-ray luminosity
obtained by Quintana \& Melnick (1982) implies X-ray luminosities in
the right range. We can therefore assume that the arc cross sections
of our simulated clusters are typical for X-ray luminous clusters in
the EMSS survey.

The curves in Fig.~\ref{fig:4.3} give $n_{\rm c}^{-1}\,\tau$. Using
the number density of bright EMSS clusters given above,
\begin{equation}
  \tau(r\ge10) \sim \left\{\begin{array}{ll}
  2.9\times10^{-6} & \hbox{(OCDM)} \\
  3.3\times10^{-7} & \hbox{($\Lambda$CDM)} \\
  4.4\times10^{-8} & \hbox{(S/$\tau$CDM)} \\
  \end{array}\right.\;.
\label{eq:6.1}
\end{equation}
Since the whole sky has $\sim4.1\times10^4$ square degrees, the total
solid angle in which sources at $z_{\rm s}\sim1$ are imaged as large
arcs with $r\ge10$ is
\begin{equation}
  \delta\omega \sim \left\{\begin{array}{ll}
  1.2\times10^{-1}\,\hbox{sq.\ deg.} & \hbox{(OCDM)} \\
  1.4\times10^{-2}\,\hbox{sq.\ deg.} & \hbox{($\Lambda$CDM)} \\
  1.8\times10^{-3}\,\hbox{sq.\ deg.} & \hbox{(S/$\tau$CDM)} \\
  \end{array}\right.\;.
\label{eq:6.2}
\end{equation}
The sources which are imaged as arcs with the above properties,
$r\ge10$ and $R\le21.5$ correspond to sources with $R\la23.5$ because
of the magnification. Taking the number densities compiled and
measured by Smail et al., there are $\sim2\times10^4$ such sources per
square degree. The average redshift of such sources is $\sim0.8-1$
(e.g.\ Lilly et al.\ 1995). Since the average arc-cluster redshift in
our models is at $z_{\rm c}\sim0.3-0.4$, the exact redshift of sources
at $z\sim0.8-1$ has only very little influence; the critical surface
mass density changes by $\sim10\%$ when sources are shifted from
$z=0.8$ to $z=1.2$. It follows that the number of such arcs on the
whole sky expected from our simulations is
\begin{equation}
  N_{\rm arcs} \sim \left\{\begin{array}{ll}
  2400 & \hbox{(OCDM)} \\
   280 & \hbox{($\Lambda$CDM)} \\
    36 & \hbox{(S/$\tau$CDM)} \\
  \end{array}\right.\;.
\label{eq:6.3}
\end{equation}
There are $\sim7500$ clusters on the sky which match the criteria of
the EMSS bright cluster sample (Le F\`evre et al.\ 1994). Taking the
number of arcs per cluster found in the EMSS clusters, the expected
number of arcs on the whole sky is $\sim1500-2300$. Despite the
obvious uncertainties in this estimate, {\em the only of our
cosmological models for which the expected number of arcs comes near
the observed number is the open CDM model\/}. The others fail by one
or two orders of magnitude. The large differences in arc optical depth
between the cosmological models investigated makes this result fairly
insensitive to moderate uncertainties. It therefore appears fair to
conclude that arc statistics in the framework of cluster-normalised
CDM models demands that $\Omega_0$ is low, and that $\Omega_\Lambda$
is small. Conversely, if $\Omega_0\la1$, clusters have to form earlier
than in our models. We estimate with the simple Press-Schechter
approach sketched above that in order to achieve that,
$\sigma_8\sim1.2-1.3$ would be necessary in the SCDM case.

We have neglected the potential influence of cD galaxies or cooling
flows on the arc cross sections that could increase the central
surface mass densities of the clusters and thus also their arc cross
sections. Most probably, this influence is small compared to the huge
differences between the cosmological models. Nonetheless, we will
study this issue in detail in a further paper because individual
cluster galaxies may well affect arc morphologies, if not their total
number.

\begin{acknowledgements} 

We have benefited from many instructive discussions, particularly
with Suvendra Dutta, Gus Evrard, Eelco van Kampen, Shude Mao, Houjun
Mo, Julio Navarro, Peter Schneider, and Simon White. Matthias
Steinmetz and Achim Weiss contributed a lot to the numerical code
which we used to automatically classify the arcs.
This work was supported in part by the Sonderforschungsbereich 375 of
the Deutsche Forschungsgemeinschaft.

\end{acknowledgements} 

\end{document}